# SURFACE GENERATION ANALYSIS IN MICRO END-MILLING CONSIDERING THE INFLUENCES OF GRAIN

*Jinsheng Wang[1,2]  Yadong Gong[1]  Garbriel Abba[2]  Kui Chen[1]  Jiashun Shi[1]  Guangqi Cai[1]*

[1]P.O. 319, Laboratory of Advanced Manufacturing and Automation,
Northeastern University, Shenyang, Liaoning, China, 110004, +86-24-83687229,
jinsheng.wang.neu@gmail.com
[2]Laboratoire de Génie Industriel et de Production Mécanique, Université de Metz - LGIPM-CEMA -
Ile du Saulcy, F-57045 METZ Cedex 01, France

## ABSTRACT

Micro end-milling method is a universal micro manufacturing method, which can be used to fabricating complex 3D structures and parts with many materials. But compared with their micrometer order size, their surface roughness quality is not satisfied. In this paper, the different metal phase grains influences are researched, and the micro end-milling process is described while the material is anisotropic. In this paper, the physical characteristics of different grains, especially friction coefficient $\mu$ and elastic module $E$ , are very critical to determine the chip formation process and surface generation. The chip is often discontinues because of the grain boundary effect. Through the micro end-milling experiment, the bottom surface results correlate very well with the theory analysis.

## 1. INTRODUCTION

Micro end-milling is a very useful processing method in many diverse fields and industries, including telecommunications, portable consumer electronics, defense, and biomedical, with the latter being perhaps the largest beneficiary from miniaturized technological enhancements. Compared with other micro manufacturing methods, especially the methods based on semi-conductor processing techniques, micro end-milling has a lot of advantages. It is suitable for accommodating individual components rather than large batch sizes, and has the ability to monitor the in-process quality of components so that problems can be corrected during fabrication. It is capable for fabricating 3D free-form surfaces, which is especially important for the production of micro-injection molds. So the micro end-milling is a very important method to compensate the disadvantage of the manufacturing methods based on Si in MEMS and MOEMS fields. Nowadays, the several bio-micro-electro-mechanical systems (bio-MEMS) facilities are currently investigating ways to produce bio-MEMS devices based on the micromechanical processes of plastic micro-injection and hot embossing.

The results of micro end-milling are in two aspects. One is the part with very small size, such as micro column, micro die and molds and micro blades. Fig.1 and Fig.2 show typical micro parts manufactured by micro milling machine [1]. The other one is some normal part with very small features, such as micro hole, micro flute and others. Fig.3 and Fig.4 are micro flute and micro spiral respectively [2].

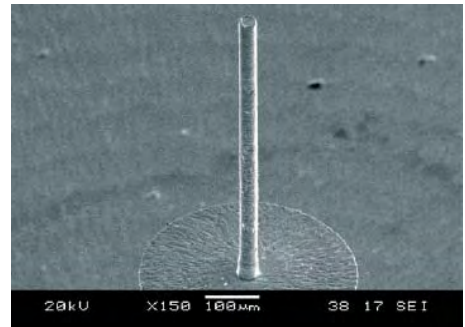

Fig.1 Micro column

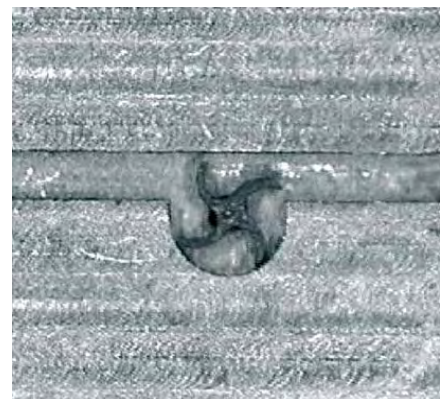

Fig.2 Micro impeller

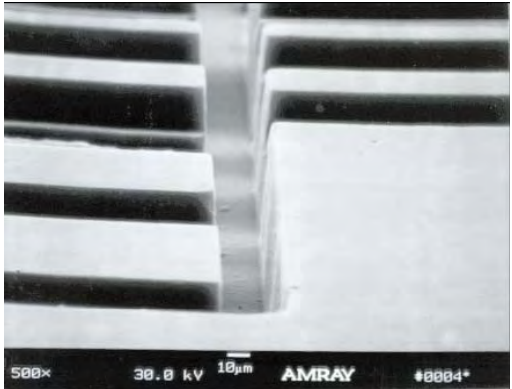

Fig.3 Micro flute

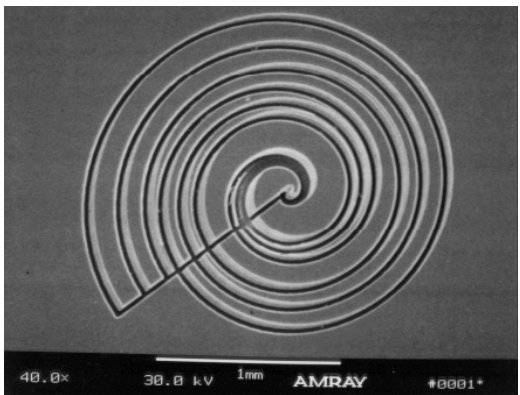

Fig.4 Micro spiral

## 2. SURFACE ROUGHNESS PROBLEM

To the conventional mechanical machined products, their dimensions are in the millimeter order, and their surface roughness is in the micrometer order according to the requirement. This surface quality could be satisfied to its usages.

To these micro products, their shapes are very delicate, which belong to the precision manufacturing fields. However, compared with their small size in micrometer order, their surface quality, which are always in sub-micrometer order, are not very satisfied. This phenomenon is caused by many factors, such as processing parameters, machine tool and materials and the main reason is the scaled size effect.

Surface generation in the micro-endmilling process was studied by Vogler et al. [3]. The surface roughness was found to be strongly affected by the tool edge radius. With the increasing of the tool edge radius, the surface roughness is increasing. In a fixed tool edge radius, the feed rate will play an important role.

S. Filiz, et al [4] measured the surface roughness of the bottom of the micro milled channels. The results indicate that the surface generation mechanism at these speeds includes not only geometric considerations, but also the minimum chip thickness, ploughing, and elastic recovery effects.

M. Takacs, et al [5] found that the harder the material the better the surface quality is. And the lifetime of the tool is better, if the material is more ductile.

From these literatures, previous investigators have done some works about the surface generation problem in some aspects, such as minimum chip thickness, feed rates, cutting speeds, and so on. Some researchers have resorted to experimentation aimed at some special materials. Therefore, it is not suitable to be applied for a wide range of materials.

In this paper, the research focuses on the influences of grains in metallic materials to the surface generation process and the machined surface roughness problem.

## 3. SURFACE GENERATION PROCESS

Compared to the conventional milling, the size of the tool and the cutting parameters in micro end-milling shrink to micrometer order, which are less than or equal to the grain size. So the metal grain influences play an important role to the micro end-milled surface generation process.

### 3.1. Micro end-milling process

In macro scale, when the tool tip is engaging the workpiece, there are lots of grains with different metal phases being cut. Statically, the influences of the material microstructures are neglected. And the macro physical characteristics of the workpiece play important roles, such as the hardness, tensile strength, elastic module and thermal conductivity. So the workpiece material is regarded as isotropic, and the surface roughness is decided mainly by the milling parameters.

To the micro end-milling, the grain sizes of the workpiece are between 0.1μm and100μm, and the edge radii of the tool tip are less than 5μm. When the tool rotates once, along the length of the tool edge, there are a few grains being cut, which are countable. Their influences are different because of their pronounced physical characteristics differences, and the workpiece is regard as anisotropic. The Fig.5 is an illustration of the cutting zone in the micro mill's tool tip, when the influences of the metal grains are considered.

### 3.2. Minimum chip thickness in one grain

Vertical to the edge, the tool tip engages into one grain, which is shown in Fig.6. At this moment, the grain is considered as isotropic. Because the cutting depth $a$ is always smaller than the edge radius $r$ , it is a large negative rake cutting process. In this process, the minimum chip thickness $h_m$ plays an important role. Before achieving $h_m$, the material is accumulated and forms a little heave. There is severe ploughing phenomenon and the material is compressed and recovered along the flank face. When $h_m$ is achieved, the chip is formed and slips along the rake face. According to the reference [6], $h_m$ is relative to two factors. The first one is the friction coefficient between workpiece and tool $\mu$, which correlates with the grain physical characteristics. The second one is the tool tip edge radius $r$. Given these two parameters, $h_m$ is calculated by Eq. (1) and Eq. (2).

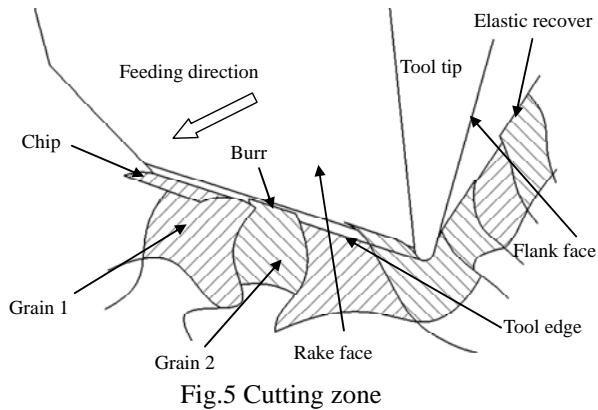

Fig.5 Cutting zone

$$\cos\beta = \frac{1}{\sqrt{1+\mu^2}} \qquad (1)$$

$$h_m = r(1-\cos(\pi/4 - \beta/2)) \qquad (2)$$

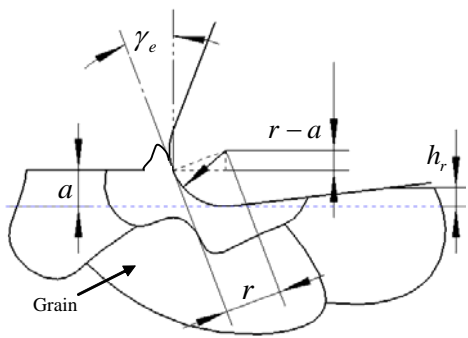

Fig.6 Orthogonal Micro Cutting

### 3.3. Grain influences

To the poly crystalline material, which are composed by different metal grains, there physical characteristics are obvious different. So in Eq.(1), the friction coefficient $\mu$ is different to them, and $h_m$ is different too. So in the micro milling process, as depicted in Fig.5, there are four grains dispersed along the cutting edge, because of their different $h_m$ the chip formation statuses are different. To some larger $\mu$ grain, the chip is formed, but to some smaller $\mu$ grain, only a little burr is formed or just being compressed.

### 3.4. Grain boundary influences

The metal atoms, which exist in the grain boundary, arrange very abnormally. And there are lots of impurities in the grain boundary. So the deform resistance of the grain boundary is bigger, and it is not easy to be removed like the normal grain. Therefore, the chip usually breaks on the boundary.

### 3.5. Elastic recover

After the tool edge passing, the grains will recover under the flank face, because of the different elastic module $E$ of these grains, the recover heights $h_r$ for them are different, which can be calculated with Eq.(3).

$$h_r = \begin{cases} h_m, \sigma < \sigma_P \\ h_m - \dfrac{\sigma_P \sqrt{r^2 - (r-h_m)^2}}{E}, \sigma \geq \sigma_P \end{cases} \qquad (3)$$

Where $\sigma$ is gotten by Eq.(4), which means the compressing stress inflicting on the grain. $\sigma_P$ is the proportional stress limit of the grain. $r$ is the tool tip edge radius.

$$\sigma = \frac{E \cdot h_m}{\sqrt{r^2 - (r-h_m)^2}} \qquad (4)$$

## 4. EXPERIMENT VALIDATION

A micro end-milling experiment is conducted and the bottom surface was measured to validate the theory before.

### 4.1. Experiment setup

The experiment is done in a three dimensional micro machine tool developed by us as Fig.7, which are driven by linear motor with 10nm resolution and 50mm stroke and equipped with a high speed spindle. A microscope is used to monitoring the milling process.

A four-flute tungsten carbide micro end-mill with 0.396mm diameter is used. The edge radius is 1.36 μm measured using SEM as Fig.8.

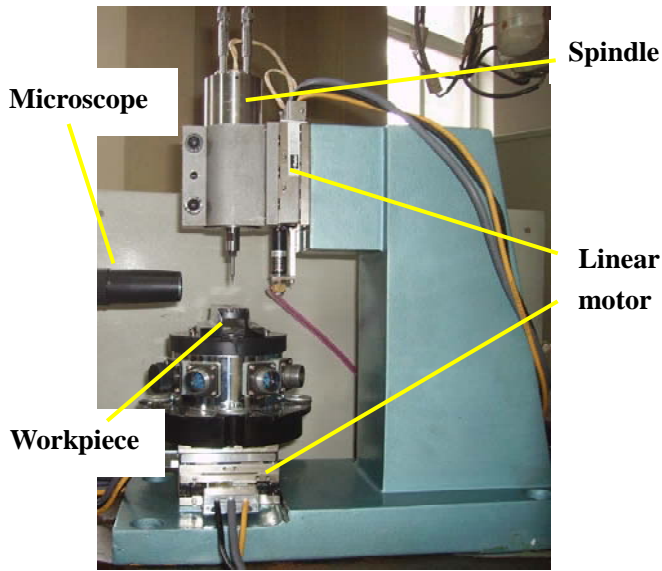

Fig.7 Micro machine tool

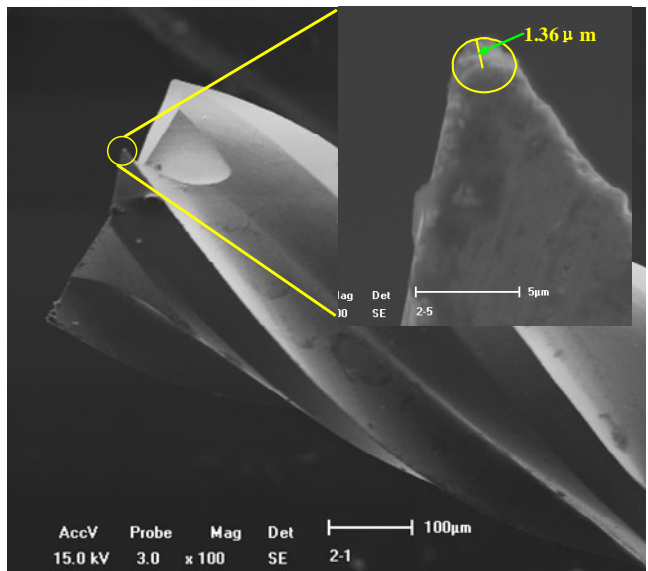

Fig.8 SEM photo of Φ0.396mm micro end-mill
with 1.36 μ m edge radius

The workpiece is Al6061 alloy, and the microstructure is as Fig.9. The white part is the soft phase, which is mainly composed by Al, and its average grain intercept length is 10.238μm. The black part is the brittle phase, which mainly is acicular silicon with 1.854μm average grain intercept length. The material characteristics and the calculation results $h_m$, $h_r$ based on Eq.(1) to Eq.(4) for the soft phase and brittle phase are listed in the Table 1. From the comparison, $h_m$ in brittle phase grain is larger than the soft phase grain, which means the chip forms easier in the brittle phase grain. $h_r$ is smaller in brittle phase grain than in soft phase grain, which means the recover height is lower in the brittle phase grain.

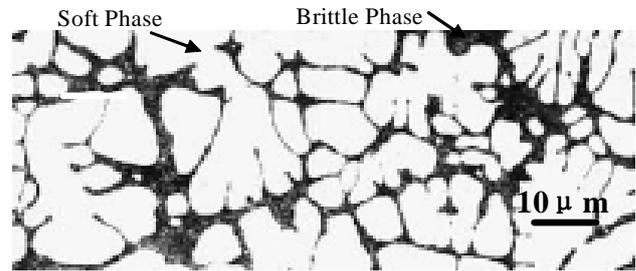

Fig.9 Microstructure of Al6061

|  | Soft phase | Brittle phase |
|---|---|---|
| $E$ [Gpa] | 70 | 8.7 |
| $\mu$ | 0.3 | 0.5 |
| $\sigma_P$ [MPa] | 240 | 0.04 |
| $h_m$ [μm] | 0.2689 | 0.2031 |
| $h_r$ [μm] | 0.2655 | 0.2021 |

Table.1 Al6061 material characteristics
and calculated results

### 4.2. Experiment results

The micro end-milled bottom surface SEM photo is as Fig.10 (a), the surface measured result using CHR 150 Stil Micromeasure 2 optical surface measurement system is as (b), and (c) is the roughness curve corresponding to the line in (b).

In the Fig.10 (a), other than the tool passing path, the pronounced phenomenon is that the concaves forms on the brittle phase grains and the chip is obvious around it, which correlate well with the description before. From (b) and (c), the large-scale curves correspond to the brittle phase. And the small-scale curves correspond to the minimum chip thickness influence. And the distances showed in (a) and (c) correlate the soft phase average intercept length well.

Fig.11 shows the chips, which on the bottom of the micro end-mill, are highly fragmented, indicating the discontinuous and interrupted nature on the grain boundaries of the chip formation process.

### 5. CONCLUSIONS

In this paper, the surface analysis is done to the micro end-milling operation, which is a very useful micro fabrication method to applying many diverse materials to the MEMS products. The main analysis factor is the different grain influences, because their different physical

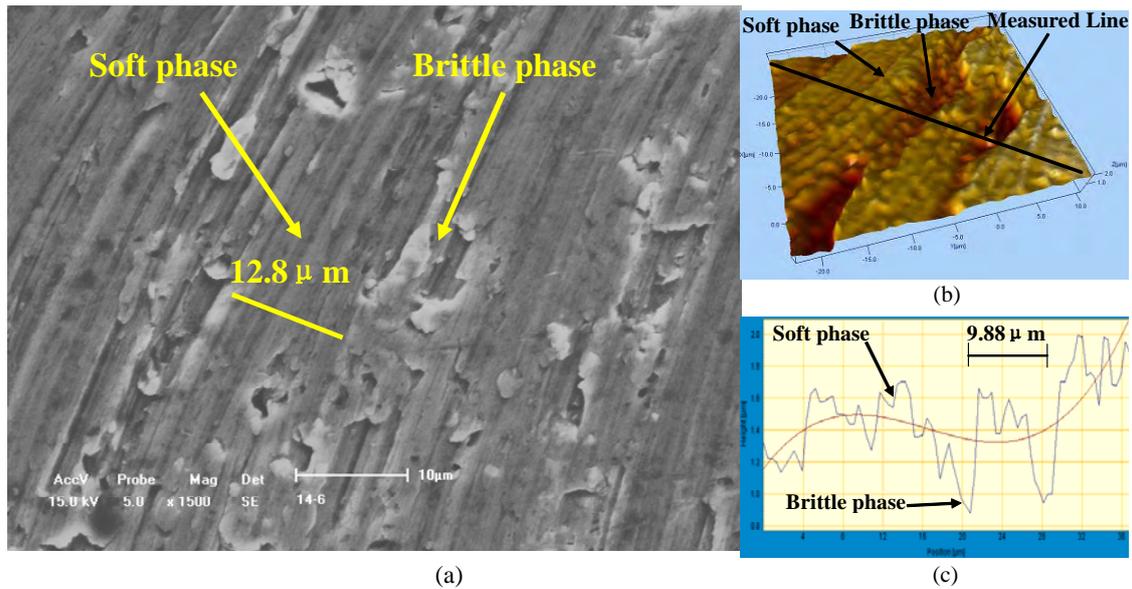

(a)  (b)  (c)

Fig.10 Micro end-milled bottom surface (Spindle speed: 25000rpm; Cutting depth: 10μm; Feed rate: 20mm/min)

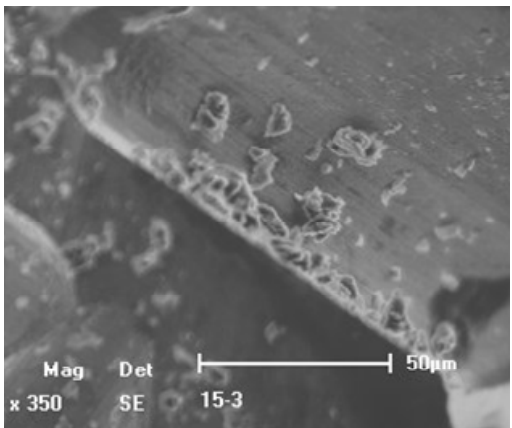

Fig.11 chips in micro end-milling

characteristics, especially $\mu$ and $E$, the minimum chip thickness, elastic recover are different to them. So the surface generation in micro end-milling process is affected by them pronouncedly. And the chip is always discontinues because the grain boundary influence. Through the micro end-milling experiment, the results correlate very well to the theory.

## ACKNOWLEGEMENT

This work is supported by the National 985-Ⅱ foundation. The authors would also like to thank Mr. Wei Wang and Mr. Jianfeng Cao for their assistance in processing the workpiece material and manufacturing the experiment equipments.